\documentclass[11pt]{article}

\usepackage[margin=1in]{geometry}
\usepackage{graphicx}
\usepackage{booktabs}
\usepackage{amsmath}
\usepackage[hyphens]{url}
\usepackage{xcolor}
\usepackage{caption}
\usepackage{natbib}
\usepackage{listings}
\usepackage{hyperref}

\graphicspath{{figures/}}

\hypersetup{
  colorlinks=true,
  linkcolor=blue,
  citecolor=blue,
  urlcolor=blue
}

\title{From Patent Expiry to Business Pathways:\\
AI Workflows for Activating Innovation Archives}

\author{
  Sidney Shapiro\\
  Dhillon School of Business, University of Lethbridge\\
  \texttt{sidney.shapiro@uleth.ca}\\
  \and
  Mark Price\\
  Opus College of Business, University of St. Thomas\\
  \texttt{pric4635@stthomas.edu}
}

\date{}

\begin{document}

\maketitle

\begin{abstract}
Patent databases represent one of the largest public archives of technical knowledge, yet much of this knowledge remains difficult to identify, interpret, and reuse once patent rights expire or lapse. This paper proposes an AI-enabled framework for discovering expired and lapsing patents, identifying technology trends, and translating patent disclosures into business pathways. We use pathways to mean structured commercialization routes such as SaaS products, services, licensing packages, consulting playbooks, training offerings, data products, or internal process tools. The framework treats patent expiry as both a business signal and an archival transition, not primarily as a legal problem. Legal status remains important, but it is one risk-screening input alongside customer need, implementation feasibility, channel access, and market timing. We describe a system architecture that combines patent metadata, maintenance-fee records, legal-status indicators, semantic search, patent-family analysis, market signals, and generative AI workflows. A proof of concept parses all 378 records in an official weekly CIPO ST.96 archive, identifies 20 expired, lapsed, or near-expiry candidates, tests the stability of the transparent scoring model, and uses a locally hosted Qwen3.6 model to populate structured review packets. The evaluation demonstrates reproducible ingestion, stable rankings under weight perturbation, and schema-conformant model output, while also exposing incomplete legal-status coverage and the need for register and expert review. We argue that AI can function as a discovery and translation layer for dormant technical knowledge, but that such systems must explicitly represent legal uncertainty, data limitations, and commercialization risk.
\end{abstract}


\section{Introduction}
\label{sec:introduction}

Patent databases are among the largest publicly accessible archives of technical knowledge. In exchange for temporary monopoly rights, inventors disclose detailed specifications that describe methods, systems, and applications across nearly every industry. Patent disclosure becomes public well before expiry in many jurisdictions, but expiry or lapse can change whether the claimed invention remains enforceable in a particular jurisdiction. In practice, this disclosed knowledge remains difficult to discover, interpret, and operationalize at scale.

The barrier is not merely access. Patent records are written in specialized legal and technical language. Legal status depends on jurisdiction-specific maintenance rules, patent families span multiple national registers, and commercial viability depends on market context that is rarely encoded in patent metadata. Entrepreneurs, university technology transfer offices, small and medium enterprises (SMEs), and corporate innovation teams therefore face a translation problem: public archives exist, but actionable business pathway maps do not.

This paper proposes an AI-enabled framework for treating expired and lapsing patents as a \emph{dynamic innovation archive}. We frame patent expiry not only as a legal status change but also as an archival transition event and a business pathway signal, a point at which disclosed technical knowledge may become more available for reuse, reinterpretation, and commercialization. We use the term \emph{pathway} to emphasize that the central task is exploration: identifying plausible routes from disclosed knowledge to business action. These routes may include SaaS products, productized services, consulting methods, training programs, licensing bundles, data products, internal workflow tools, or research commercialization projects. Our contribution combines three themes:

\begin{enumerate}
  \item \textbf{Archival intelligence:} patents as public innovation infrastructure rather than static legal repositories;
  \item \textbf{AI-mediated translation:} NLP and workflow automation that convert patent disclosures into ranked business pathways and product hypotheses;
  \item \textbf{Responsible commercialization support:} explicit representation of legal uncertainty, family risk, customer need, implementation feasibility, and pathway-specific assumptions.
\end{enumerate}

We adopt a design-science orientation \citep{hevner2004design}, presenting a conceptual architecture, Python workflow patterns, and evaluation-oriented artifacts rather than a production patent search platform. The workflows emphasize auditable decision logic for translating archival records into pathway hypotheses without implying verified legal status or commercial viability for any individual patent.

Our research questions are:
\begin{enumerate}
  \item How can expired and lapsing patents be represented as an actionable innovation archive?
  \item What AI workflows are needed to translate patent disclosures into product, service, SaaS, data, training, or licensing pathways?
  \item What business, legal, ethical, and archival risks emerge when public patent knowledge is operationalized for commercialization?
  \item How can pathway-scoring models balance technical feasibility, customer value, market access, and intellectual-property risk?
\end{enumerate}

The remainder of the paper proceeds as follows. Section~\ref{sec:background} reviews patent systems and data access. Section~\ref{sec:related} surveys related work in patent NLP and expired-patent opportunity research. Section~\ref{sec:system} presents the system architecture. Section~\ref{sec:workflows} describes Python-backed workflows. Sections~\ref{sec:implications}--\ref{sec:limitations} discuss implications and limitations. Section~\ref{sec:conclusion} concludes.

\section{Background}
\label{sec:background}

\subsection{Patent Systems and Expiry Events}

Patents grant inventors time-limited exclusive rights in exchange for public disclosure. In the United States and Canada, utility patents generally expire twenty years from the earliest effective filing date, subject to maintenance or annuity fee requirements. Failure to pay required fees can cause patents to lapse before nominal term expiration. These events, including expiry, lapse, and abandonment, are legally distinct and must not be conflated in analytics pipelines.

Patent families link related filings across jurisdictions through priority claims, continuations, and divisionals. A patent may be expired in one country while an equivalent family member remains active elsewhere. International analytics therefore require both national register reconciliation and cross-border family mapping via services such as EPO Open Patent Services (OPS) and INPADOC legal events \citep{epo2024ops,wipo2024patentscope}.

\subsection{Patents as Innovation Archives}

Archival theory emphasizes that records acquire meaning through context, provenance, and use \citep{yeo2007concepts,saa2024provenance}. Patent databases traditionally function as retrieval systems for legal and technical documents. We argue they should also be understood as \emph{innovation archives} whose value increases when expiry signals are linked to interpretive metadata: legal status, family relationships, semantic similarity, implementation difficulty, and commercialization readiness.

Expiry is therefore not merely a legal endpoint. It is an archival and business signal indicating that disclosed knowledge may transition from protected asset to reusable technical input, subject to family risk, regulatory constraints, customer demand, implementation cost, and market conditions.

\subsection{Data Access: United States, Canada, and Beyond}

Patent analytics systems depend on tiered data ingestion because no single source provides authoritative legal status, full text, and family linkage globally.

\textbf{United States.} The USPTO Open Data Portal (ODP) provides bulk data, patent file-wrapper records, maintenance-fee event data, and API-based search \citep{uspto2024odp,uspto2026patentsview}. Maintenance fee payment status can be verified through the USPTO Maintenance Fees Storefront. PatentsView and Google Patents Public Data on BigQuery offer research-friendly longitudinal tables and full text \citep{google2024patents}. Recent API migrations highlight the need for robust sync architectures that tolerate source churn.

\textbf{Canada.} The Canadian Intellectual Property Office (CIPO) distributes patents primarily through IP Horizons bulk downloads, ST.96 XML files, and quarterly researcher CSV/TXT datasets \citep{cipo2024iphorizons,cipo2026maintenance}. Unlike the US, Canada does not offer a public patent REST API; programmatic access is bulk-first. Records may appear in English, French, or both, adding NLP complexity. Legal status events conform to WIPO ST.27 when provided in ST.96 exports.

\textbf{International context.} EPO OPS, WIPO PATENTSCOPE, PATSTAT, and Lens.org provide family linkage, legal events, and aggregated bibliographic data across jurisdictions \citep{lens2024api}. Other national offices, including JPO, KIPO, and CNIPA, maintain distinct access models. Production systems should treat international coverage as optional enrichment layers rather than assuming uniform API availability.

Table~\ref{tab:us-canada-access} summarizes US--Canada differences most relevant to our architecture.

\begin{table}[t]
\centering
\caption{Comparison of patent data access in the United States and Canada.}
\label{tab:us-canada-access}
\small
\begin{tabular}{p{2.2cm}p{5.2cm}p{5.2cm}}
\toprule
\textbf{Dimension} & \textbf{United States} & \textbf{Canada} \\
\midrule
API access & USPTO Open Data Portal (ODP) APIs; PatentsView migration & No public patent REST API; bulk/XML downloads \\
Full text & ODP file wrappers; BigQuery public dataset & ST.96 XML; quarterly researcher CSV/TXT \\
Expiry drivers & 20-year term; maintenance fees at 3.5, 7.5, 11.5 years & 20-year term; annual annuities from second anniversary \\
Authoritative lapse check & Maintenance Fees Storefront & Admin status in register; fee history \\
Family/legal status & ODP plus EPO OPS/INPADOC & ST.27 in ST.96; EPO OPS for foreign equivalents \\
Language & English & English and French bilingual records \\
Update cadence & Daily/weekly ODP deltas & Weekly XML; quarterly researcher sets \\
\bottomrule
\end{tabular}
\end{table}

\subsection{NLP for Patent Documents}

Patent NLP differs from general-domain text processing because claims, the legally operative portion of a patent, use specialized syntax, antecedent basis, and long coordination structures \citep{verberne2016claim}. Recent surveys document widespread use of transformer models for classification, retrieval, and summarization \citep{sun2024nlp,shomee2025patentanalysis}. Domain-adapted models such as PatentBERT demonstrate that claims alone can support CPC classification tasks \citep{lee2019patentbert}.

For entrepreneurship workflows, however, text-only models are insufficient. Pathway identification requires integrating claims and descriptions with maintenance status, family risk, citation networks, customer segments, implementation assumptions, route-to-market choices, and scoring rubrics that encode commercial feasibility and legal uncertainty.

\section{Related Work}
\label{sec:related}

\subsection{Patent NLP and Analytics}

Patent analytics has long supported R\&D intelligence, competitive landscaping, and technology forecasting \citep{bloom2013patents}. Earlier technology-mining work treated patents as structured traces of inventive activity that could be combined with bibliometrics, assignee networks, and classification codes to identify emerging domains \citep{porter2005techmining}. Patent landscaping practice similarly emphasizes search strategy, corpus curation, family grouping, and analyst interpretation rather than document retrieval alone \citep{wipo2015landscape}. These traditions are important for expired-patent analytics because opportunity discovery depends on the quality of the constructed corpus before any AI model is applied.

NLP methods address classification, clustering, citation analysis, summarization, and semantic retrieval. Transformer-based approaches now dominate, with task-specific adaptations for multi-label CPC prediction, claim parsing, and similarity search \citep{sun2024nlp,lee2019patentbert}. Recent surveys also note a shift from single-document classification toward multimodal and workflow-oriented patent analysis, where text, drawings, citations, legal events, and external market signals are combined \citep{shomee2025patentanalysis}. Our framework follows this direction by treating AI outputs as intermediate artifacts in a larger analytic process.

Claim parsing remains challenging because dependent claims inherit limitations from parent claims, and antecedent references must be resolved before scope can be interpreted \citep{verberne2016claim}. These challenges matter for product translation workflows: a plain-language summary that ignores claim dependency structure may overstate or understate protectable scope.

\subsection{Technology Opportunity Discovery}

Technology opportunity discovery uses patent corpora to identify promising directions for R\&D, product development, and diversification. Studies in this area often combine semantic similarity, citation links, technology life-cycle indicators, and market or product terms to surface candidate opportunities \citep{song2017tod}. The expired-patent setting changes the problem: the question is not only whether a technology area appears promising, but which business pathways might make specific disclosed knowledge useful. Legal status matters, but the pathway question is broader because the same disclosure can support different business routes with different customers, revenue models, capabilities, and risks.

This distinction creates a need for richer pathway representations. A ranked patent list may be useful for expert searchers, but entrepreneurs and SMEs need explanations, route-to-market options, implementation assumptions, and warnings about what the ranking does not prove. Our workflow design therefore emphasizes interpretable scores, pathway hypotheses, and review packets rather than optimization of retrieval metrics alone.

\subsection{Expired Patents and Open Innovation}

A growing literature treats expired patents as sources of reusable technical knowledge. \citet{park2021expired} analyze expired patents as technology opportunities, emphasizing that public-domain disclosures can reduce search costs for firms seeking implementable ideas. Work on higher-education institutions similarly frames expired patents as resources for academic entrepreneurship \citep{frontiers2023expired}.

This literature connects naturally to work on absorptive capacity and open innovation. Firms benefit from external knowledge only when they can recognize, assimilate, and apply it \citep{cohen1990absorptive}. Open innovation similarly argues that useful knowledge often lies outside firm boundaries and must be actively searched, evaluated, and recombined \citep{chesbrough2003open}. Expired patents can therefore be viewed as an underused external knowledge pool, but only for organizations with enough technical, legal, and market capability to interpret them.

Our contribution extends this line by integrating expiry analytics with archival framing, multi-jurisdiction data engineering, and AI workflows that produce structured pathway hypotheses and review packets rather than retrieval results alone. The aim is not to automate invention or legal clearance. It is to reduce the interpretive burden that prevents public patent knowledge from becoming practically usable.

\subsection{Decision Support and Legal-Tech Boundaries}

Patent analytics tools often blur the boundary between information retrieval and legal advice. Freedom-to-operate (FTO) analysis requires expert judgment across active family members, continuations, trademarks, regulatory requirements, and product-specific claim mapping. Credible systems must therefore function as \emph{decision support}: they organize evidence, flag uncertainty, and route users to professional review rather than declaring legal safety.

This boundary is especially important for generative AI systems because fluent summaries can obscure uncertainty. Risk-management guidance for AI systems emphasizes context, documentation, transparency, and human oversight for consequential decisions \citep{nist2023airmf}. In patent opportunity workflows, those principles translate into provenance fields, confidence annotations, visible legal cautions, and exportable records that experts can inspect.

Design-science research in information systems provides a suitable methodological anchor for presenting an artifact-oriented framework with explicit evaluation criteria \citep{hevner2004design}. Rather than claiming a universal model of commercialization success, the present paper specifies an artifact class, its intended users, its data dependencies, and the evaluation evidence that future deployments should collect.

\section{System Design}
\label{sec:system}

\begin{figure}[t]
  \centering
  \includegraphics[width=\linewidth]{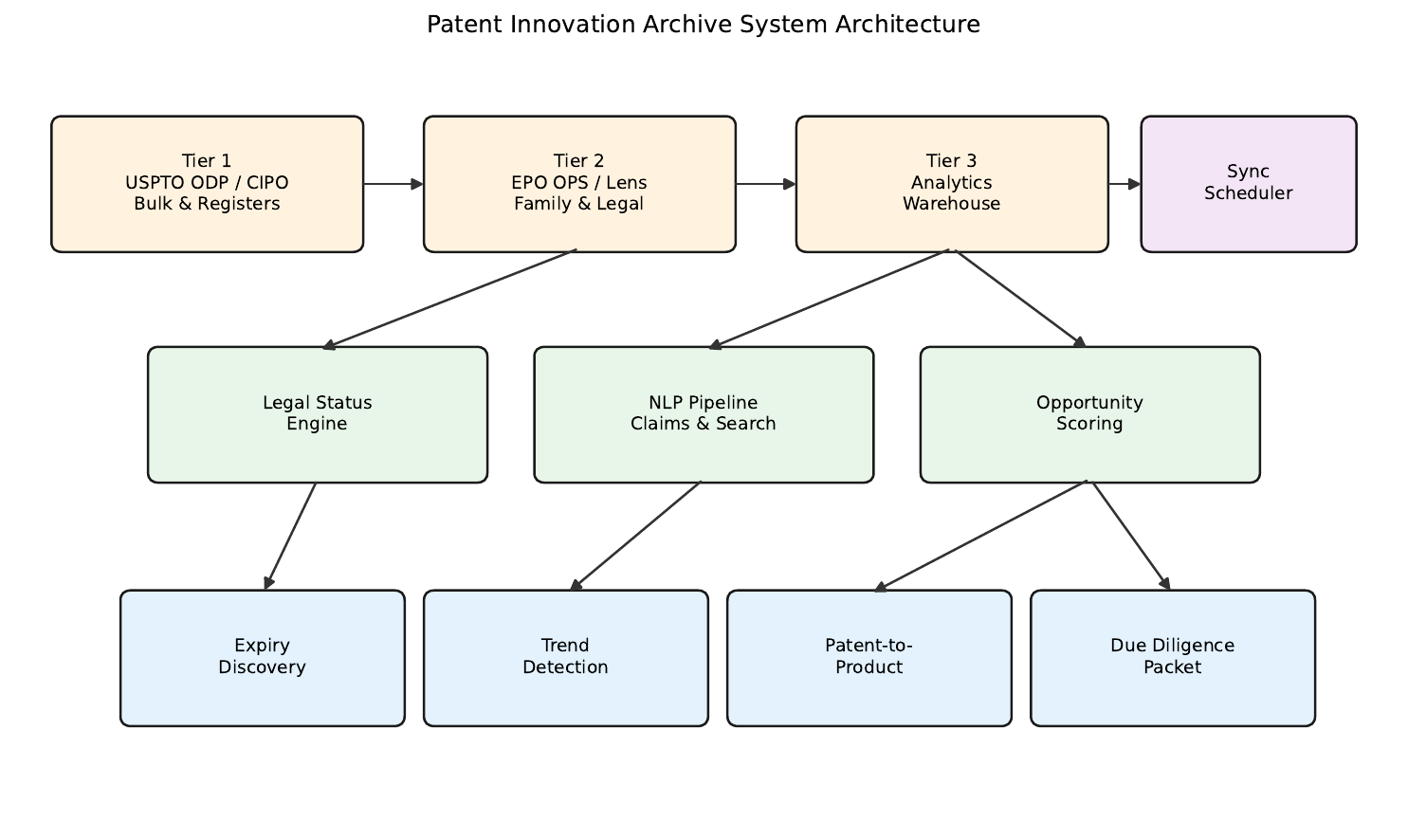}
  \caption{Patent innovation archive system architecture with tiered ingestion, NLP processing, and workflow outputs.}
  \label{fig:architecture}
\end{figure}

We propose a tiered architecture that separates authoritative national sources, cross-border aggregators, and derived analytics stores. Figure~\ref{fig:architecture} illustrates the major components.

\subsection{Tiered Data Ingestion}

\textbf{Tier 1 (national authority).} USPTO ODP file wrapper and maintenance-fee records; CIPO ST.96 XML and researcher datasets for Canadian patents. Tier 1 sources are preferred for legal-status reconciliation.

\textbf{Tier 2 (aggregators).} EPO OPS for family and legal events; Lens.org or BigQuery for cross-jurisdiction search and enrichment; WIPO PATENTSCOPE for PCT collections.

\textbf{Tier 3 (analytics warehouse).} A canonical schema stores normalized patent identifiers, bibliographic fields, legal events, family identifiers, embeddings, opportunity scores, and workflow outputs. Sync jobs combine bulk baseline loads with incremental deltas and on-demand enrichment for candidate patents.

\subsection{Database and Table Design}

The system should not be implemented as a Python script operating over loose files. Python is useful for ingestion, scoring, NLP calls, and packet generation, but the durable analytic object is a database. A production version should use a relational database or analytical warehouse, such as PostgreSQL, DuckDB, BigQuery, or Snowflake, with tables that preserve source provenance and support repeatable joins across patent, legal-status, family, market, compliance, and contract data.

Listing~\ref{lst:schema} sketches a minimum relational schema. The intent is not to prescribe one vendor-specific database, but to show the normalized entities that the workflows require.

\begin{lstlisting}[language=SQL,caption={Minimum warehouse schema for patent pathway analytics.},label={lst:schema}]
CREATE TABLE patents (
    patent_id TEXT PRIMARY KEY,
    jurisdiction TEXT NOT NULL,
    publication_number TEXT,
    application_number TEXT,
    title TEXT,
    abstract TEXT,
    claims_text TEXT,
    filing_date DATE,
    grant_date DATE,
    estimated_expiry_date DATE,
    source_url TEXT,
    source_snapshot_id TEXT,
    updated_at TIMESTAMP
);

CREATE TABLE legal_events (
    event_id TEXT PRIMARY KEY,
    patent_id TEXT REFERENCES patents(patent_id),
    event_date DATE,
    event_code TEXT,
    event_text TEXT,
    source_system TEXT,
    confidence NUMERIC
);

CREATE TABLE patent_families (
    family_id TEXT,
    patent_id TEXT REFERENCES patents(patent_id),
    jurisdiction TEXT,
    legal_status TEXT,
    PRIMARY KEY (family_id, patent_id)
);

CREATE TABLE cpc_codes (
    patent_id TEXT REFERENCES patents(patent_id),
    cpc_code TEXT,
    PRIMARY KEY (patent_id, cpc_code)
);

CREATE TABLE pathway_scores (
    patent_id TEXT REFERENCES patents(patent_id),
    run_id TEXT,
    opportunity_score NUMERIC,
    commercial_relevance NUMERIC,
    saas_feasibility NUMERIC,
    family_risk TEXT,
    score_explanation JSON,
    PRIMARY KEY (patent_id, run_id)
);

CREATE TABLE pathway_packets (
    packet_id TEXT PRIMARY KEY,
    patent_id TEXT REFERENCES patents(patent_id),
    run_id TEXT,
    candidate_pathways JSON,
    evidence_bundle JSON,
    unresolved_risks JSON,
    reviewer_decision TEXT,
    created_at TIMESTAMP
);
\end{lstlisting}

Adjacent public-record tables can follow the same pattern. Procurement notices, compliance filings, standards references, contracts, and litigation records should each have stable identifiers, source URLs, collection timestamps, normalized text fields, and embedding references. The important design rule is to keep raw source snapshots, normalized fields, derived features, and generated interpretations separate. This makes it possible to reproduce a pathway packet, inspect a model output, rerun a scoring method, or explain why a record appeared in a candidate list.

\subsection{Legal-Status Engine}

The legal-status engine computes estimated expiry dates, maintenance states, and family-risk indicators. It applies jurisdiction-specific rules: US maintenance windows at 3.5, 7.5, and 11.5 years; Canadian annuities from the second anniversary. When sources disagree, national register status overrides derived estimates. All outputs include provenance fields and confidence annotations.

\subsection{NLP Pipeline}

The NLP pipeline performs:
\begin{enumerate}
  \item document segmentation (claims, description, abstract);
  \item normalization and OCR cleanup for legacy scans;
  \item CPC/IPC classification for domain filtering;
  \item dense retrieval over claims and abstracts for semantic domain search;
  \item generative summarization and pathway-hypothesis drafting with structured output schemas;
  \item scoring models for commercial relevance, pathway fit, and SaaS feasibility.
\end{enumerate}

Claims-first retrieval aligns with patent NLP findings \citep{lee2019patentbert}, while generative layers must be constrained by structured metadata to reduce hallucination risk. In the proposed pipeline, AI is used for retrieval, normalization, classification, summarization, and pathway drafting, but not for unreviewed legal clearance. Deterministic rules reconcile dates and register events; NLP models translate dense technical and legal text into reviewable claims, assumptions, and questions.

The decision-making layer is therefore hybrid. Embedding models retrieve related patents and external records; classifiers route documents into domains and pathway types; extraction models identify claim elements, actors, inputs, outputs, and operational constraints; generative models draft summaries and pathway hypotheses under a schema. The system then asks a narrower decision question: which candidates deserve expert attention, what evidence supports that attention, and what uncertainty must be resolved before action?

\subsection{Opportunity Scoring}

Pathway scores combine commercial relevance, SaaS feasibility, expiry or lapse status, and family-risk penalties. Scores are explanatory rather than predictive in the illustrative implementation: each component is exposed to users and auditors. This design supports entrepreneurship use cases where transparency matters as much as ranking accuracy. A high score should be read as a prompt for pathway exploration, not as a finding that a business is viable or legally cleared.

\subsection{Workflow Orchestration}

Four workflow modules consume the analytics warehouse:
\begin{itemize}
  \item expiry discovery;
  \item trend detection;
  \item patent-to-pathway translation;
  \item pathway review packet generation.
\end{itemize}

\begin{figure}[t]
  \centering
  \includegraphics[width=\linewidth]{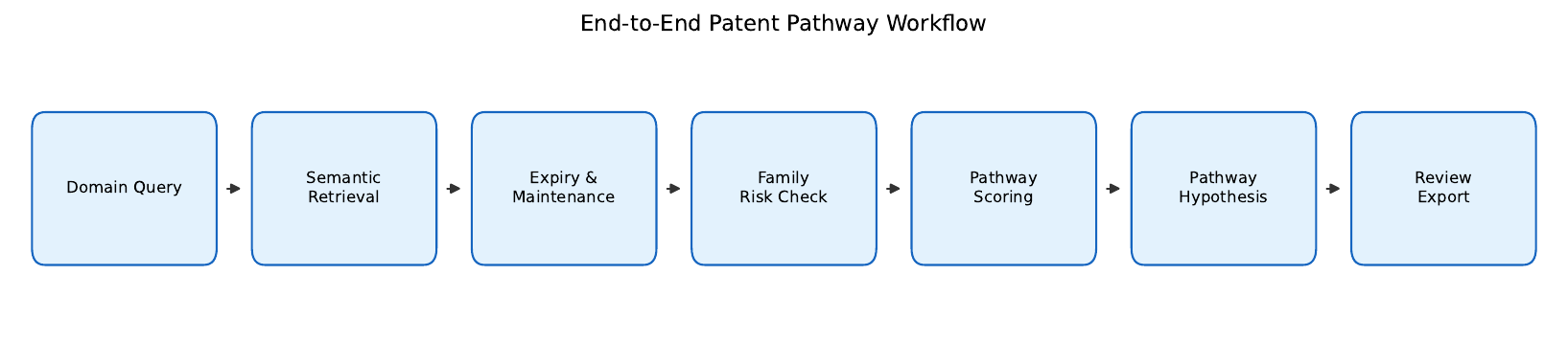}
  \caption{End-to-end workflow from domain query to pathway review export.}
  \label{fig:workflow-pipeline}
\end{figure}

Figure~\ref{fig:workflow-pipeline} shows the end-to-end path from domain query to exportable expert-review packet.

The modules pass structured objects rather than static reports. Expiry discovery returns candidate records with status provenance and family risk. Trend detection returns clusters and abandonment signals that can be joined to market, compliance, or contract evidence. Patent-to-pathway translation returns candidate business routes with assumptions and required inputs. The review-packet generator binds these outputs into a decision artifact that preserves source evidence, model outputs, uncertainty fields, and recommended next checks. Table~\ref{tab:workflow-outputs} summarizes the inputs, processing steps, and primary outputs of each module.

\begin{table}[t]
\centering
\caption{Workflow inputs, processing steps, and outputs.}
\label{tab:workflow-outputs}
\small
\begin{tabular}{p{2.4cm}p{4.8cm}p{4.8cm}}
\toprule
\textbf{Workflow} & \textbf{Processing Steps} & \textbf{Primary Outputs} \\
\midrule
Expiry discovery & Semantic domain filter; expiry/lapse check; maintenance review; family risk flag & Ranked opportunity list; plain-language summaries; scores \\
Trend detection & Time-series filing analysis; CPC clustering; assignee abandonment signals & White-space clusters; domain signals; structured review prompts \\
Patent-to-pathway & Claim summarization; pathway hypothesis generation; MVP scoping & Product, service, SaaS, training, and internal-tool concepts \\
Pathway review & Register reconciliation; family mapping; citation/continuation checks; market assumptions & Structured review packet; uncertainty notes; expert actions \\
\bottomrule
\end{tabular}
\end{table}

\section{Python Workflow Patterns}
\label{sec:workflows}

We describe four Python workflow patterns for converting patent records into reviewable pathway artifacts. These examples should be read as orchestration code over the database schema in Listing~\ref{lst:schema}, not as a recommendation to build the system in Python alone. In practice, SQL or warehouse queries retrieve candidate records, Python applies scoring, NLP, and packet-generation logic, and the resulting pathway packets are written back to durable tables for review and audit. The purpose of the code is to make the decision logic auditable: what data are required, where AI is used, which checks remain deterministic, and which judgments must be deferred to human reviewers or counsel.

\subsection{Data Model}

The canonical \texttt{PatentRecord} dataclass is a typed application-layer view of rows from \texttt{patents}, \texttt{patent\_families}, \texttt{cpc\_codes}, and \texttt{pathway\_scores}. It captures bibliographic fields, legal status, family identifiers, scoring attributes, and text excerpts. Derived methods include \texttt{days\_to\_expiry}, \texttt{family\_risk}, \texttt{opportunity\_score}, and \texttt{score\_explanation}. Listing~\ref{lst:data-model} is deliberately close to runnable Python: a reader could hydrate it from SQL query results, CSV, JSON, ODP, CIPO, or OPS and then reuse the workflow functions below.

\begin{lstlisting}[language=Python,caption={Usable patent record model and scoring helpers.},label={lst:data-model}]
from dataclasses import dataclass
from datetime import date
from enum import Enum

class LegalStatus(str, Enum):
    ACTIVE = "active"
    EXPIRED = "expired"
    LAPSED = "lapsed"
    UNKNOWN = "unknown"

@dataclass(frozen=True)
class PatentRecord:
    patent_id: str
    title: str
    abstract: str
    claims_excerpt: str
    jurisdiction: str
    filing_date: date
    expiry_date: date | None
    legal_status: LegalStatus
    maintenance_status: str
    cpc_codes: list[str]
    family_id: str
    active_family_members: int
    source_url: str
    commercial_relevance: float = 0.5
    saas_feasibility: float = 0.5

    def days_to_expiry(self, as_of: date) -> int | None:
        if self.expiry_date is None:
            return None
        return (self.expiry_date - as_of).days

    @property
    def family_risk(self) -> str:
        if self.active_family_members >= 3:
            return "high"
        if self.active_family_members > 0:
            return "medium"
        return "low"

    def opportunity_score(self, as_of: date) -> float:
        status_bonus = 0.15 if self.legal_status in {
            LegalStatus.EXPIRED, LegalStatus.LAPSED} else 0.0
        days = self.days_to_expiry(as_of)
        expiry_bonus = 0.10 if days is not None and 0 <= days <= 730 else 0.0
        family_penalty = min(0.35, self.active_family_members * 0.08)
        raw = (0.40 * self.commercial_relevance
               + 0.30 * self.saas_feasibility
               + status_bonus + expiry_bonus - family_penalty)
        return round(max(0.0, min(1.0, raw)), 3)

    def score_explanation(self, as_of: date) -> dict[str, object]:
        return {
            "score": self.opportunity_score(as_of),
            "status": self.legal_status.value,
            "days_to_expiry": self.days_to_expiry(as_of),
            "family_risk": self.family_risk,
            "active_family_members": self.active_family_members,
            "source": self.source_url,
        }
\end{lstlisting}

The weights in Listing~\ref{lst:data-model} are heuristic and intentionally transparent. In a deployed system, they should be calibrated against expert labels or observed downstream decisions. The implementation should expose each component so that analysts can perform sensitivity checks rather than treating the score as a black-box prediction.

\subsection{Workflow 1: Expiring Patent Discovery}

The expiry discovery workflow filters patents by domain keywords, selects records that are expired or nearing expiry within a configurable horizon, and ranks candidates by opportunity score. Outputs include plain-language summaries, source links, legal-status provenance, and structured metadata for each candidate. Listing~\ref{lst:expiry} shows a complete filter that can operate over any iterable of \texttt{PatentRecord} objects.

\begin{lstlisting}[language=Python,caption={Expiry discovery filter and ranking.},label={lst:expiry}]
def domain_match(patent: PatentRecord, keywords: list[str]) -> bool:
    text = " ".join([
        patent.title, patent.abstract, patent.claims_excerpt,
        " ".join(patent.cpc_codes)
    ]).lower()
    return any(k.lower() in text for k in keywords)

def discover_candidates(
    patents: list[PatentRecord],
    keywords: list[str],
    as_of: date,
    max_days: int = 730,
    threshold: float = 0.45,
) -> list[dict[str, object]]:
    rows = []
    for patent in patents:
        if not domain_match(patent, keywords):
            continue
        days = patent.days_to_expiry(as_of)
        expired = patent.legal_status in {
            LegalStatus.EXPIRED, LegalStatus.LAPSED}
        nearing = days is not None and 0 <= days <= max_days
        if not (expired or nearing):
            continue

        score = patent.opportunity_score(as_of)
        if score < threshold:
            continue

        rows.append({
            "patent_id": patent.patent_id,
            "title": patent.title,
            "score": score,
            "status": patent.legal_status.value,
            "days_to_expiry": days,
            "family_risk": patent.family_risk,
            "why_ranked": patent.score_explanation(as_of),
            "source_url": patent.source_url,
        })
    return sorted(rows, key=lambda r: r["score"], reverse=True)
\end{lstlisting}

\subsection{Workflow 2: Trend Detection}

Trend detection aggregates filing counts by year, counts expired patents by CPC prefix, and identifies assignees abandoning patent families. In this paper, trend analysis is treated as an input to pathway review rather than as a stand-alone chart. The useful output is not a line graph; it is a set of candidate technology clusters with enough context to decide whether further review is warranted.

Listing~\ref{lst:trend} shows the core pattern. The workflow treats each CPC prefix as a rough technology bucket, compares expired and active coverage, and emits a white-space signal only when expired records outnumber active records and unresolved family risk remains low. The output is a structured cluster list, not a chart, so it can feed directly into pathway review.

\begin{lstlisting}[language=Python,caption={Trend and white-space detection over CPC prefixes.},label={lst:trend}]
from collections import Counter, defaultdict

def cpc_prefix(code: str) -> str:
    return code.split("/")[0]

def whitespace_clusters(
    patents: list[PatentRecord],
    min_expired: int = 3,
    max_active: int = 1,
) -> list[dict[str, object]]:
    expired_by_cpc = Counter()
    active_by_cpc = Counter()
    examples = defaultdict(list)

    for patent in patents:
        for code in patent.cpc_codes:
            prefix = cpc_prefix(code)
            if patent.legal_status in {
                LegalStatus.EXPIRED, LegalStatus.LAPSED}:
                expired_by_cpc[prefix] += 1
                examples[prefix].append(patent.patent_id)
            elif patent.legal_status == LegalStatus.ACTIVE:
                active_by_cpc[prefix] += 1

    clusters = []
    for prefix, expired_count in expired_by_cpc.items():
        active_count = active_by_cpc[prefix]
        if expired_count >= min_expired and active_count <= max_active:
            clusters.append({
                "cpc_prefix": prefix,
                "expired_count": expired_count,
                "active_count": active_count,
                "example_patents": examples[prefix][:5],
                "review_prompt": (
                    "Check whether demand, compliance pressure, "
                    "or implementation cost has changed.")
            })
    return sorted(clusters, key=lambda c: c["expired_count"], reverse=True)
\end{lstlisting}

The same pattern can be extended beyond filing counts. A practical deployment should join patent records with maintenance-fee events, family-level legal events, assignee histories, product categories, public procurement notices, compliance filings, standards references, incident reports, or contract-tracking data. These adjacent datasets help distinguish a technically interesting disclosure from an actionable pathway. For example, an expired safety-monitoring disclosure is more meaningful if procurement contracts show continuing demand for inspection software, compliance filings reveal recurring reporting burdens, or standards changes create new documentation requirements.

\subsection{Workflow 3: Patent-to-Pathway Translation}

Given a selected patent, the translation workflow produces a \texttt{ProductHypothesis} object that is used as a pathway hypothesis. It includes a technical summary, plain-language claim explanation, product interpretations, SaaS options, required data inputs, user segments, MVP features, competitive substitutes, legal caution flags, and recommended next steps. The workflow explicitly treats legal review as one pathway gate, not as the whole business problem.

Listing~\ref{lst:product} illustrates the structured-output approach. In a production system, the lists would be generated or enriched by a constrained LLM call, but the output schema would remain explicit so downstream tools can validate required fields. The example below uses a small rule catalog because it is transparent and testable; a model can draft the prose, but the pathway categories, required evidence, and caution flags remain controlled.

\begin{lstlisting}[language=Python,caption={Patent-to-pathway structured output.},label={lst:product}]
PATHWAY_RULES = {
    "saas": {
        "keywords": ["dashboard", "alert", "monitor", "workflow"],
        "evidence": ["buyer demand", "integration data", "security review"],
    },
    "service": {
        "keywords": ["inspection", "audit", "assessment"],
        "evidence": ["repeatable method", "expert delivery capacity"],
    },
    "training": {
        "keywords": ["procedure", "checklist", "supervisor"],
        "evidence": ["curriculum fit", "regulatory training need"],
    },
    "compliance_tool": {
        "keywords": ["report", "record", "incident", "compliance"],
        "evidence": ["filing burden", "forms", "retention obligations"],
    },
}

def translate_to_pathways(patent: PatentRecord) -> dict[str, object]:
    text = f"{patent.title} {patent.abstract} {patent.claims_excerpt}".lower()
    caution = [
        "Verify legal status against the national register.",
        "Review active family members before commercialization.",
    ]
    if patent.active_family_members:
        caution.append(
            f"{patent.active_family_members} active family member(s) remain.")

    options = []
    for name, rule in PATHWAY_RULES.items():
        hits = [k for k in rule["keywords"] if k in text]
        if hits:
            options.append({
                "pathway": name,
                "match_terms": hits,
                "required_evidence": rule["evidence"],
                "first_test": f"Interview 5 target users about {name}.",
            })

    return {
        "patent_id": patent.patent_id,
        "technical_summary": patent.abstract,
        "plain_language_claims": (
            f"This record describes {patent.title.lower()}. "
            f"Key claim text: {patent.claims_excerpt[:500]}"),
        "candidate_pathways": options,
        "legal_caution_flags": caution,
    }
\end{lstlisting}

The pathway object should be treated as a branching device, not as a recommendation. A single disclosure can produce several candidate routes, each with its own evidence requirements. A SaaS pathway requires data integration, workflow adoption, cybersecurity review, and subscription willingness. A consulting or training pathway may require less software development but more domain credibility and repeatable delivery methods. A contract-tracking or compliance pathway depends on whether the disclosed method maps to forms, audits, obligations, and renewal events that organizations already manage. Expanding the code around pathways is therefore more useful than showing a small synthetic table of ranked records.

\subsection{Workflow 4: Pathway Review Packet}

The pathway review workflow assembles a structured packet for expert review. It verifies expiry and maintenance indicators from sample metadata, reports family and citation risk tiers, notes international equivalent status, and lists recommended actions including formal FTO review when needed. The packet states clearly that it is not legal advice, but its main purpose is broader: organizing business, technical, and legal questions so a team can decide which pathway deserves further exploration.

Listing~\ref{lst:diligence} shows how legal uncertainty is represented as review metadata rather than as a binary clearance decision. The function builds an exportable packet that can be serialized as JSON, attached to a CRM record, sent to a legal reviewer, or used as an agenda for customer discovery.

\begin{lstlisting}[language=Python,caption={Pathway review packet construction.},label={lst:diligence}]
def build_review_packet(
    patent: PatentRecord,
    hypothesis: dict[str, object],
    evidence: dict[str, object],
    as_of: date,
) -> dict[str, object]:
    unresolved = [
        "This packet supports pathway review and is not legal advice.",
        "National register status should override derived estimates.",
    ]
    if patent.family_risk != "low":
        unresolved.append("Active family members require counsel review.")
    if patent.legal_status == LegalStatus.UNKNOWN:
        unresolved.append("Legal status is unknown or not yet reconciled.")

    return {
        "patent_id": patent.patent_id,
        "generated_on": as_of.isoformat(),
        "status": patent.legal_status.value,
        "maintenance_status": patent.maintenance_status,
        "family_risk": patent.family_risk,
        "score": patent.opportunity_score(as_of),
        "hypothesis": hypothesis,
        "evidence": evidence,
        "unresolved_risks": unresolved,
        "next_actions": [
            "Confirm legal status in the national register.",
            "Review family members through OPS or INPADOC.",
            "Select one pathway for customer discovery.",
            "Estimate build, delivery, and compliance costs.",
            "Commission FTO review if launch risk warrants it.",
        ],
    }
\end{lstlisting}

Listing~\ref{lst:evidence} sketches the additional evidence-gathering layer. The goal is to convert archival research into action by attaching each pathway hypothesis to the records that would support or weaken it. In practice, each \texttt{search\_*} function can wrap a real API, bulk-file query, or manually curated spreadsheet.

\begin{lstlisting}[language=Python,caption={Evidence hooks for pathway assessment.},label={lst:evidence}]
def collect_evidence(patent: PatentRecord, clients: dict) -> dict[str, object]:
    terms = [patent.title, patent.abstract, *patent.cpc_codes]
    evidence = {
        "register": clients["register"].status(patent.patent_id),
        "family": clients["family"].events(patent.family_id),
        "contracts": clients["contracts"].search(terms, limit=10),
        "compliance": clients["compliance"].search(
            terms=["inspection", "incident", "audit", *terms],
            limit=10),
        "standards": clients["standards"].match(patent.claims_excerpt),
    }
    evidence["review_questions"] = [
        "Do contracts show recurring buyer demand?",
        "Do compliance records show a repeated reporting burden?",
        "Do standards or regulations make adoption more likely?",
        "Does any active family member cover the same pathway?",
    ]
    return evidence

def make_packet(patent: PatentRecord, clients: dict, as_of: date) -> dict:
    hypothesis = translate_to_pathways(patent)
    evidence = collect_evidence(patent, clients)
    return build_review_packet(patent, hypothesis, evidence, as_of)
\end{lstlisting}

\subsection{Real-Data Proof of Concept}

To move beyond synthetic examples, we implemented a bounded proof of concept using an official CIPO ST.96 weekly bulk archive. The workflow downloads the July 2025 weekly CIPO XML package, parses \emph{every} patent record it contains (378 records in this release, rather than the 25-record sample used in an earlier draft), and for each record extracts title, abstract, claims excerpt, assignee, filing date, publication date, document status, and IPC classifications. It estimates a twenty-year expiry date from the filing date, derives an estimated legal status, computes the expiry-focused opportunity score from Section~\ref{sec:workflows}, and writes a structured JSON review artifact together with a provenance manifest that records the archive URL, its SHA256 hash, and the parse date.

The score now applies the same expiry and status bonuses used in Listing~\ref{lst:data-model}, so expired, lapsed, and near-expiry records are ranked as expiry-discovery candidates rather than being crowded out by high-interest active filings. Of the 378 parsed records, 11 are estimated expired or lapsed and 20 fall within the expiry-discovery window (expired, lapsed, or within 730 days of estimated expiry).

\begin{table}[t]
\centering
\caption{Expiry discovery over the full weekly CIPO ST.96 archive. The five records shown are the top expired, lapsed, or near-expiry candidates by opportunity score. Days to expiry is estimated from the twenty-year term; a lapsed or inactive status can arise earlier through abandonment, so those rows may still show a positive term estimate. The final row is the highest-scoring active record, included as a contrast: high pathway interest does not imply public-domain readiness.}
\label{tab:real-cipo-poc}
\footnotesize
\begin{tabular}{p{1.7cm}p{6.5cm}p{2.2cm}r r}
\toprule
\textbf{Record} & \textbf{Title} & \textbf{Est. status} & \textbf{Days to expiry} & \textbf{Score} \\
\midrule
CA2616010 & Medication Compliance System And Associated Methods & active or pending & 11 & 0.44 \\
CA3129424 & Peelable Coloured Paint Swatches & lapsed or inactive & 4,961 & 0.41 \\
CA2680705 & Materials For And Method For Manufacturing Container With Stacking Shoulders And Resulting Container & lapsed or inactive & 1,174 & 0.41 \\
CA2687626 & Glass Yarns Suitable For Reinforcing Organic And/Or Inorganic Materials & lapsed or inactive & 676 & 0.40 \\
CA3097433 & Method For Treatment Of Elements Obtained By An Additive Manufacturing Process & lapsed or inactive & 4,664 & 0.34 \\
\midrule
CA3090703 & Configuration Systems And Methods For Secure Operation Of Networked Transducers & active or pending & 4,631 & 0.54 \\
\bottomrule
\end{tabular}
\end{table}

Table~\ref{tab:real-cipo-poc} shows the top five expiry-discovery candidates, with the highest-scoring active record appended as a contrast. The purpose of the run is not to claim that these records are commercially attractive or legally available for reuse. The contrast row makes the central caution concrete: an active record can outscore genuine near-expiry records on pathway interest alone, so the workflow must separate pathway interest from public-domain readiness. The proof of concept therefore validates the ingestion, scoring, and packet-generation pattern over the full archive, not the business value of any individual record.

For the local drafting layer we used the Qwen3.6 model served through a local Ollama instance, with sampling temperature 0.1, JSON-constrained output, and hybrid reasoning disabled so that the token budget produces the requested structured fields rather than hidden chain-of-thought. The model drafted a technical summary, candidate pathways, required evidence, legal cautions, and reviewer questions for the top near-expiry record and for the active contrast record; a deterministic template supplies the same fields when the model is unavailable. Appendix~\ref{app:ollama} reproduces one full prompt and response. In our trial, model summaries were useful for quick technical orientation, but pathway evidence and reviewer questions still benefited from deterministic schema controls. This supports the design choice made throughout the framework: model outputs should be routed into structured packets with source fields, uncertainty labels, and human-review gates rather than treated as final recommendations.

We selected a local model for reasons that are operational as well as technical. Patent records are public, but pathway packets may combine them with proprietary customer notes, unpublished product concepts, internal scoring assumptions, or counsel-directed review questions. Local inference keeps these inputs within the analyst's environment, avoids transmitting them to a third-party model provider, and permits organizations to apply their own retention, access-control, and audit policies. It also fixes the model version and decoding settings for reproducibility, allows offline or restricted-network deployment, and makes per-record processing costs more predictable at larger volumes. These benefits do not make local output inherently more accurate; the same provenance, schema validation, uncertainty labels, and human-review requirements apply regardless of where a model is hosted.

\subsection{Evaluation}
\label{sec:evaluation}

Because the framework is an artifact rather than a validated product, we report bounded, automatic evaluation over the real and synthetic corpora. These measures characterize ingestion coverage, status-derivation consistency, scoring robustness, and structured-output quality. They are not a substitute for expert inter-rater assessment of pathway quality, which we identify as future work in Section~\ref{sec:limitations}.

\textbf{Corpus composition.} Table~\ref{tab:eval-corpus-stats} summarizes the parsed CIPO corpus. Most records are active or pending, which is expected for a recent weekly archive, and only a small minority are expired or lapsed. This distribution is itself a finding: a single weekly release is a weak source of public-domain candidates, so production expiry discovery should accumulate multiple releases or query historical baselines rather than a single week.

\textbf{Status-derivation check.} Table~\ref{tab:eval-status-check} maps the raw CIPO document-status text to the derived category. The keyword rule classifies granted, examination, and abandonment states confidently, but leaves compliant and allowed applications as \emph{unknown}, giving definitive coverage of roughly 77\%. The residual \emph{unknown} bucket reflects incomplete vocabulary coverage, not evidence of expiry, and marks a concrete target for improving the legal-status engine.

\textbf{Scoring sensitivity.} Table~\ref{tab:eval-sensitivity} perturbs each transparent weight by $\pm 20\%$ and reports top-5 overlap and Kendall's $\tau$ against the baseline ranking. Rankings are stable ($\tau$ above 0.87 on the real corpus and above 0.90 on the synthetic corpus, with top-5 overlap of four or five out of five), which indicates that the heuristic weights are explanatory and robust rather than finely tuned to a particular result.

\textbf{Structured-output quality.} Table~\ref{tab:eval-ollama} records the Qwen3.6 drafting-layer behavior. Every live call returned the five required keys and parsed as valid JSON without repair, confirming that a constrained local model can populate the packet schema reliably. This is an output-format check, not a judgment of summary fidelity, which requires expert review.

\textbf{Illustrative ranking.} To show the expiry-focused score behaving as intended when public-domain records are present, Table~\ref{tab:sample-results} reports the top of the synthetic workplace-safety corpus, where expired records with low family risk rise to the top of the ranking. The synthetic corpus complements the real archive: the real data validates ingestion and scoring at scale, while the synthetic data illustrates the expiry-discovery ranking that a multi-release production corpus would surface.

\begin{table}[t]
\centering
\caption{Real CIPO ST.96 corpus composition. The proof of concept now parses the entire weekly archive (378 records of 378 XML entries) rather than a 25-record sample. Status is derived from the CIPO document-status text and a twenty-year term estimate as of the paper reference date.}
\label{tab:eval-corpus-stats}
\small
\begin{tabular}{lrr}
\toprule
\textbf{Estimated status} & \textbf{Records} & \textbf{Share} \\
\midrule
estimated expired & 3 & 0.8\% \\
lapsed or inactive & 8 & 2.1\% \\
active or pending & 279 & 73.8\% \\
unknown & 88 & 23.3\% \\
\midrule
\textbf{Total parsed} & 378 & 100.0\% \\
Near-expiry or expired & 20 & 5.3\% \\
\bottomrule
\end{tabular}

\vspace{0.4em}
{\footnotesize\emph{Source:} cipo\_st96\_weekly\_20250727.zip; SHA256 73c10795647018a5\ldots; reference date 2026-07-09.}
\end{table}

\begin{table}[t]
\centering
\caption{Status-derivation check: raw CIPO document-status text mapped to the derived category. The keyword rule confidently classifies granted, examination, and abandonment states, but leaves compliant and allowed applications as \emph{unknown}, so definitive coverage is 76.7\%. The residual \emph{unknown} bucket is a data-quality limitation, not evidence of expiry.}
\label{tab:eval-status-check}
\small
\begin{tabular}{p{7.0cm}p{3.2cm}r}
\toprule
\textbf{Raw CIPO status text} & \textbf{Derived category} & \textbf{Records} \\
\midrule
Examination & active or pending & 184 \\
Granted and Issued & active or pending & 81 \\
Application Compliant & unknown & 72 \\
Pre-Grant & active or pending & 14 \\
Allowed & unknown & 14 \\
Deemed Abandoned and beyond the Period of Reinstatement & lapsed or inactive & 5 \\
Granted and Issued & estimated expired & 3 \\
Deemed Abandoned & lapsed or inactive & 2 \\
Deemed Expired & lapsed or inactive & 1 \\
Conditionally Allowed & unknown & 1 \\
Entered National Phase & unknown & 1 \\
\bottomrule
\end{tabular}
\end{table}

\begin{table}[t]
\centering
\caption{Scoring sensitivity to weight perturbation. Each row perturbs one transparent weight from Listing~\ref{lst:data-model} by $\pm 20\%$ and reports top-5 overlap and Kendall's $\tau$ rank correlation against the baseline ranking, on the real CIPO corpus and the synthetic corpus. High $\tau$ values indicate the ranking is stable and the weights are explanatory rather than finely tuned.}
\label{tab:eval-sensitivity}
\small
\begin{tabular}{p{5.4cm}cccc}
\toprule
& \multicolumn{2}{c}{\textbf{Real CIPO}} & \multicolumn{2}{c}{\textbf{Synthetic}} \\
\cmidrule(lr){2-3}\cmidrule(lr){4-5}
\textbf{Perturbation} & Top-5 & $\tau$ & Top-5 & $\tau$ \\
\midrule
Commercial weight $+20\%$ & 5/5 & 0.886 & 5/5 & 0.962 \\
Commercial weight $-20\%$ & 5/5 & 0.873 & 5/5 & 0.962 \\
SaaS weight $+20\%$ & 5/5 & 0.877 & 5/5 & 0.981 \\
SaaS weight $-20\%$ & 4/5 & 0.876 & 5/5 & 0.962 \\
Status bonus $+20\%$ & 5/5 & 0.887 & 5/5 & 0.924 \\
Expiry bonus $+20\%$ & 4/5 & 0.887 & 4/5 & 0.905 \\
\bottomrule
\end{tabular}
\end{table}

\begin{table}[t]
\centering
\caption{Structured-output quality for the local Qwen3.6 drafting layer. Reasoning is disabled so the token budget produces the requested JSON rather than hidden chain-of-thought. Every live call returned the five required keys (technical summary, candidate pathways, required evidence, legal cautions, and reviewer questions) and parsed without repair.}
\label{tab:eval-ollama}
\small
\begin{tabular}{p{6.0cm}p{6.0cm}}
\toprule
\textbf{Property} & \textbf{Value} \\
\midrule
Model & qwen3.6:latest \\
Reasoning mode & disabled (\texttt{think=false}) \\
Sampling temperature & 0.1 \\
Output format & JSON (\texttt{format=json}) \\
Live structured calls & 2 \\
All required keys present & 2/2 \\
JSON parses without repair & yes \\
Example response tokens & 514 \\
Example generation time & 98.0\,s \\
\bottomrule
\end{tabular}
\end{table}

\begin{table}[t]
\centering
\caption{Top-ranked synthetic records from the illustrative workplace safety corpus.}
\label{tab:sample-results}
\small
\begin{tabular}{lllcccc}
\toprule
\textbf{Record ID} & \textbf{Juris.} & \textbf{Status} & \textbf{Family Risk} & \textbf{Comm.} & \textbf{SaaS} & \textbf{Opp.} \\
\midrule
SYN-US-007 & US & expired & low & 0.77 & 0.92 & 0.78 \\
SYN-US-005 & US & expired & low & 0.80 & 0.87 & 0.77 \\
SYN-US-002 & US & expired & low & 0.76 & 0.88 & 0.76 \\
SYN-US-010 & US & active & low & 0.88 & 0.93 & 0.73 \\
SYN-US-004 & US & active & low & 0.85 & 0.96 & 0.73 \\
\bottomrule
\end{tabular}
\end{table}

\subsection{Reproducibility}

Workflow scripts should be versioned with fixed source snapshots, schema definitions, and model prompts. A reproducible package should record the patent data release, register query time, embedding model, prompt template, scoring weights, and reviewer decisions. The practical implementation pattern is to keep deterministic code, model prompts, and human review decisions in separate files: code computes fields and packets, prompts draft summaries and questions, and reviewers record acceptance, rejection, or escalation. This separation keeps the manuscript focused on the framework while documenting how live deployments would connect to ODP, CIPO bulk feeds, OPS enrichment endpoints, compliance repositories, and contract databases.

Concretely, the proof of concept records its own provenance. A generated manifest pins the CIPO archive URL, its SHA256 hash, the number of XML entries seen and parsed, and the reference date used for expiry estimation. The full parsed corpus, the selected candidate output, the evaluation tables, and one complete Qwen3.6 request and response are written to the repository so the artifacts can be inspected without rerunning the model. The pipeline regenerates end to end through the workflow, evaluation, and paper build steps, and a deterministic mode reproduces every field except the model-drafted prose when no local model is available.

\section{Discussion and Implications}
\label{sec:implications}

\subsection{Analytics and Innovation Management}

Patent expiry analytics can support corporate white-space analysis, university technology scouting, and incubator deal flow screening. By ranking expired and lapsing patents with explicit family-risk annotations, the framework helps analysts prioritize review time rather than treating all public-domain disclosures as equally usable. The framing in this paper is deliberately pathway-oriented: the goal is to identify plausible business routes, not to solve a legal problem in isolation.

Trend detection adds temporal context: clusters of expiring patents in a domain may indicate technology areas where early patents no longer block implementation but market demand has grown, such as workplace safety analytics benefiting from modern cloud, IoT, and NLP infrastructure unavailable at original filing dates.

The framework also changes the role of patent analytics from retrospective intelligence to operational discovery. Conventional patent dashboards often ask what competitors are filing, which assignees dominate a field, or where citation activity is concentrated. Expiry-oriented workflows ask a different question: which disclosures are becoming easier to act on, and what business pathways could they support? This shift makes legal-status data, family coverage, customer fit, route to market, and implementation metadata first-order analytic variables.

For innovation managers, this suggests a staged workflow. First, domain filtering and semantic search identify a broad opportunity pool. Second, pathway classification translates each disclosure into candidate routes such as software, services, training, data, licensing, or internal operations. Third, legal-status and family-risk checks remove or downgrade records that are not ready for reuse. Fourth, market screening produces hypotheses that can be tested through customer discovery, prototyping, or expert review. The value of AI is strongest in the transitions between these stages, where large volumes of technical text must be converted into concise, reviewable artifacts.

Scaling this workflow from one weekly archive to national or international collections requires separating inexpensive corpus-wide computation from selective, higher-cost interpretation. Bulk and incremental feeds can be normalized into partitioned warehouse tables, while deterministic expiry rules, status mappings, classification filters, and precomputed embeddings reduce millions of records to a much smaller candidate set. Family reconciliation and market-evidence joins can then operate on that set, with local or hosted language models invoked only for the highest-priority records and batched where infrastructure permits. Caching source snapshots, embeddings, and generated packets prevents unnecessary recomputation, while run identifiers and versioned scoring rules preserve auditability across updates. This funnel architecture makes processing cost grow mainly with newly ingested or shortlisted records rather than requiring every patent to be reinterpreted on every run.

This also reframes patent work as one instance of a broader class of archival research that leads to action. Organizations already make decisions from other archives: compliance filings, inspection records, public procurement notices, contract renewals, standards documents, litigation dockets, grant databases, and regulatory submissions. These records are often public or semi-public, but they are fragmented, technical, and difficult to convert into operational choices. Patent archives have the same problem. Their value increases when they are connected to adjacent evidence about obligations, buyers, renewal cycles, implementation constraints, and market demand.

\subsection{Examples of Business Pathways}

A single technical disclosure may support multiple pathways. In workplace safety analytics, an expired mobile inspection patent-like record could support a SaaS pathway: a subscription platform for inspections, audit trails, corrective actions, and dashboards. The same record could support a service pathway: a consulting firm uses the disclosed workflow to deliver standardized safety audits. It could support a training pathway: the disclosure becomes the basis for supervisor checklists, simulations, and compliance workshops. It could also support an internal-tool pathway: a manufacturer adapts the concept for its own safety operations without selling software externally.

Other pathway examples differ by asset type. A sensor-monitoring disclosure may lead to an IoT integration kit, a managed monitoring service, or a data product that benchmarks safety conditions across facilities. A semantic-analysis disclosure for incident reports may lead to a root-cause analytics SaaS, an API embedded in existing EHS platforms, or an expert-review workflow for insurers and regulators. A legacy notification disclosure may not justify a standalone venture but may still support a low-cost emergency communication template for SMEs.

These examples show why the paper does not treat the problem as purely legal. Legal status helps determine whether a pathway is worth deeper review, but it does not choose the customer, pricing model, implementation stack, channel partner, or evidence needed to validate demand. The pathway frame keeps the analytic focus on business action while preserving legal caution as a necessary constraint.

The public-domain question is central but not sufficient. A patent may be expired, abandoned, or otherwise no longer enforceable in one jurisdiction, while related family members, later improvements, trademarks, copyright, trade secrets, regulatory obligations, or contractual restrictions still matter. AI can help assess whether intellectual property appears to have moved into the public domain by reconciling expiry dates, maintenance events, terminal disclaimers, family records, and national-register status. It should express that result as a provenance-backed risk assessment, not as an unconditional permission statement.

\subsection{Entrepreneurship and SME Access}

Entrepreneurs and SMEs often lack resources for comprehensive patent landscaping and FTO review. An AI-enabled archive workflow lowers search and interpretation costs by generating plain-language summaries and pathway-oriented business hypotheses. However, equitable access requires deliberate design: if only well-resourced actors can operationalize public-domain knowledge, patent archives may reinforce rather than reduce innovation inequality.

Tools intended for broad entrepreneurial use should expose scoring rubrics, provide exportable pathway packets for advisors and legal counsel, and avoid implying that expired status equals commercial freedom.

The strongest entrepreneurial use case is early-stage pathway screening rather than final go or no-go decisions. A founder can use the system to learn what technical approaches have already been disclosed, identify older solutions that may be newly practical, and prepare a more focused conversation with engineers, customers, advisors, or counsel. This can reduce wasted search effort, but it does not remove the need for market validation. Many expired patents describe technically coherent inventions that failed because the timing, cost structure, regulatory context, channel access, or customer demand was unfavorable.

The framework can also support teaching and incubator programs. Students and early-stage teams can work from structured packets that include a technical summary, candidate pathway interpretations, risk flags, and follow-up questions. This turns patent archives into learning objects for opportunity recognition, while preserving the warning that a disclosure is not the same as a viable venture.

\subsection{Institutional and Public-Sector Use}

Universities, public research organizations, and economic-development agencies have a distinct interest in dormant technical knowledge. Technology-transfer offices usually focus on owned inventions, active licensing portfolios, and sponsored research. Expired-patent archives widen the search space to include public technical knowledge that can inform new research projects, student ventures, regional innovation programs, or applied training.

Public-sector use also raises a policy question. Patent systems are justified partly by the bargain between disclosure and time-limited exclusivity. If disclosed knowledge remains practically inaccessible after expiry, the public side of that bargain is weakened. AI-mediated archive tools can help realize the disclosure function of the patent system by improving findability, comprehension, and reuse. This is an archival and information-access argument, not only an entrepreneurship argument.

The same logic applies to compliance and contract archives. A public agency or university might use NLP to identify recurring compliance burdens, compare them with expired technical disclosures, and identify where a low-cost workflow tool, training product, or shared service could reduce administrative friction. A regional development office might combine procurement data with expired patents to find technologies for which public buyers already reveal demand. In these settings, the output is not simply a search result; it is a pathway packet that turns archival traces into a testable intervention.

\subsection{Archival Practice}

Archivists and information scientists may view this framework as extending archival description into \emph{actionable intelligence}. Richer metadata, including legal status, family links, semantic clusters, and implementation difficulty, could improve reuse of public technical records. Expiry events become cataloging signals analogous to declassification or rights expiration in other archival domains.

This framing implies that patent archives should be described not only by bibliographic metadata and legal classifications, but also by reuse conditions. Provenance, jurisdiction, maintenance status, family relationships, and uncertainty fields become part of the descriptive apparatus. Such metadata can help users understand why a record appears in an opportunity list and what evidence supports its status.

At the same time, archival enrichment must avoid overclaiming. A system that labels a record as an opportunity is performing interpretation, not neutral cataloging. Interfaces and exported packets should therefore preserve the distinction between source metadata, derived analytics, and generated interpretation. This separation supports auditability and allows expert users to challenge or revise the system's assumptions.

\subsection{Responsible AI Considerations}

Generative components must be constrained by structured outputs and provenance tracking. Summaries should cite source fields; uncertainty should appear in user interfaces and exports; human review should be mandatory before commercial decisions. These principles align with responsible AI expectations in high-stakes decision support contexts.

AI is already well suited to the pipeline tasks that make this kind of assessment possible. Retrieval models can connect a technical claim to semantically similar patents, procurement descriptions, compliance obligations, and product categories even when the vocabulary differs. Extraction models can identify claim elements, deadlines, parties, jurisdictions, maintenance events, and contractual obligations. Generative models can then draft plain-language explanations, pathway hypotheses, and reviewer questions. The value comes from chaining these steps under a controlled schema so that each generated conclusion remains tied to source evidence.

For NLP-assisted decision-making, the important design choice is to separate candidate generation from decision authority. The system may recommend that a disclosure be reviewed for public-domain reuse, that a contract cluster suggests buyer demand, or that a compliance filing points to a recurring workflow problem. It should not decide, by itself, that a product is legally clear, commercially viable, or ethically appropriate. A defensible pipeline gives analysts ranked candidates, reasons, contrary signals, and unresolved questions.

Responsible design also requires attention to failure modes. A model may hallucinate product features, simplify claim scope incorrectly, ignore active family members, or produce overconfident rankings from incomplete legal-status data. These errors are not merely technical defects because they can influence investment, product strategy, or legal exposure. The safest design pattern is to make the system conservative by default: show uncertainty, require source links, penalize unresolved family risk, and avoid binary clearance labels.

Finally, the framework should be evaluated with the people who would use it. Patent attorneys, technology-transfer professionals, entrepreneurs, and archivists will judge usefulness differently. Attorneys may prioritize provenance and legal caution, while entrepreneurs may prioritize speed and product clarity. A mature system should support these different review modes without hiding the underlying evidence.

\section{Risks and Limitations}
\label{sec:limitations}

Table~\ref{tab:risk-limitations} summarizes major risk categories. We expand on the most critical limitations below.

\begin{table}[t]
\centering
\caption{Risk categories and limitations for AI-enabled patent archive workflows.}
\label{tab:risk-limitations}
\small
\begin{tabular}{p{2.2cm}p{9.6cm}}
\toprule
\textbf{Category} & \textbf{Representative Limitations} \\
\midrule
Legal & Expired status does not guarantee freedom to operate; family members may remain active; PTA/PTE adjustments affect expiry \\
Data & API churn and rate limits; source disagreement; register lag; Canada bulk-only access; scraping brittleness \\
NLP & Claim parsing difficulty; hallucination in summaries; US/EPO-heavy training bias; bilingual CA records \\
Ethical & Public-domain knowledge may benefit well-resourced actors; requires explicit uncertainty and human review \\
\bottomrule
\end{tabular}
\end{table}

\subsection{Legal Uncertainty as a Pathway Constraint}

Expired or lapsed status in one jurisdiction does not guarantee freedom to operate. Active family members, continuations, reissues, design patents, trademarks, trade secrets, and regulatory approvals may still affect commercialization. US expiry estimation is complicated by patent term adjustment (PTA), patent term extension (PTE), and terminal disclaimers. INPADOC and aggregator legal events are indicative rather than authoritative for non-EPO authorities.

Our framework therefore supports pathway review and due diligence; it does not replace counsel. The point is to make legal uncertainty visible early enough that teams can choose, revise, or abandon pathways before investing heavily.

\subsection{Data Engineering Constraints}

Patent data ecosystems experience API churn, rate limits, and migration events, as exemplified by USPTO's PEDS-to-ODP transition and intermittent PatentsView API availability. Canada's bulk-only access model requires scheduled ingestion rather than interactive scraping of HTML registers. Cross-jurisdiction schema differences (ST.36, ST.96, USPTO JSON) increase normalization cost.

Remote database sync strategies should prefer licensed bulk and API paths over brittle scraping of national office websites or commercial front ends.

\subsection{Generative AI and Public Data Access}

Generative AI introduces a new data-access paradox for public-interest analytics. The same models that make large archives easier to search and interpret have also increased the commercial value of bulk text, code, images, and platform data. In response, some data holders have tightened APIs, raised prices, restricted scraping, or moved access into licensing arrangements. Twitter/X's API changes disrupted academic and public-interest research access \citep{murtfeldt2024riptwitter}, and Reddit's API changes were widely interpreted as part of a broader shift toward monetizing platform data used by AI developers \citep{verge2023redditapi}. Recent research on web scraping similarly notes that generative AI has contributed to platforms restricting official data channels, pushing researchers toward more legally and ethically complex collection methods \citep{brown2024webscraping}.

This matters for patent and innovation-archive research because the relevant data are often public in purpose but not frictionless in practice. Patent records, maintenance events, procurement notices, compliance filings, standards references, and contract records may be intended for public inspection, yet they depend on APIs, bulk files, search portals, rate limits, robots policies, and institutional budgets. If data providers respond to AI scraping by closing endpoints or imposing broad anti-bulk restrictions, legitimate analysis of public records becomes harder even when the underlying records remain nominally public.

The risk is not only technical inconvenience. Reduced programmatic access can privilege large firms that can pay for licensed data while excluding universities, SMEs, journalists, public agencies, and independent researchers. It can also weaken reproducibility because researchers may be unable to re-run a query, archive a source snapshot, or document why a record was unavailable at a later date. For the framework proposed here, this means that data-access governance is part of the research design: systems should prefer official bulk releases, respect published terms, cache source snapshots where permitted, record access dates and endpoint versions, and distinguish unavailable data from negative evidence.

The policy implication is that public data infrastructure should separate harmful bulk extraction from accountable public-interest analysis. Anti-scraping rules may be necessary to protect privacy, server capacity, creator rights, and platform sustainability, but overly broad restrictions can undermine the public value of records that were disclosed precisely so they could be examined. For patent archives, the disclosure bargain is weakened if AI-driven access controls make public technical knowledge practically inaccessible to the smaller organizations most likely to need search and translation support.

\subsection{NLP and Generative AI Limits}

Claim language remains out-of-domain for general LLMs because patent claims are not ordinary technical prose. They use dependencies, antecedent references, nested limitations, jurisdiction-specific conventions, and intentionally broad drafting strategies. A model that produces a fluent summary may still miss the operative limitation, collapse dependent claims into independent ones, or describe the specification as if every disclosed embodiment were actually claimed. This matters because pathway assessment depends on the difference between what a document teaches, what it claims, what has expired, and what may still be protected elsewhere.

The appropriate use of NLP is therefore staged. First, retrieval and classification models can reduce search cost by finding related patents, assigning likely technology domains, and connecting patent text to adjacent records such as procurement descriptions, compliance filings, standards, or contracts. Second, extraction models can identify entities, claim elements, dates, jurisdictions, assignees, maintenance events, family members, and technical inputs or outputs. Third, generative models can translate these structured inputs into plain-language summaries, pathway hypotheses, reviewer questions, and caution flags. At each stage, the output should remain linked to source fields so that a reviewer can inspect why the model produced a candidate pathway.

The reason for this staged design is that different pipeline tasks carry different risks. Semantic retrieval errors may omit relevant prior art or overemphasize superficially similar documents. Extraction errors may misread maintenance events, miss an active family member, or attach a legal event to the wrong jurisdiction. Generative errors may invent product features, overstate public-domain status, or turn an uncertain signal into a confident recommendation. These failures are especially consequential when the output influences investment, product design, or freedom-to-operate review.

Bilingual Canadian records require language-aware pipelines that can preserve English and French source text, avoid mixing claim terms across languages, and handle cases where abstracts, claims, or legal events are unevenly translated. Training data bias toward US and EPO corpora may also underperform on smaller jurisdictions, older scanned records, or administrative event codes that are poorly represented in general model training. For this reason, production systems should evaluate models by jurisdiction and document type rather than assuming that strong performance on US patent abstracts generalizes to Canadian legal-status reconciliation or pathway assessment.

Generative AI should not be used as an unbounded adviser in this context. It should operate inside schemas that require citations to source fields, explicit uncertainty labels, missing-data indicators, and separation between source metadata, deterministic legal-status calculations, model-generated interpretation, and human review. The practical goal is not to automate a final decision about whether intellectual property is in the public domain. The goal is to make a complex archive more navigable by producing a concise, evidence-linked packet that helps experts decide what to verify next and why.

\subsection{Evaluation Scope}

The present paper develops architecture, reproducible workflow patterns, and a bounded CIPO ST.96 proof of concept rather than a validated production system. The proof of concept now parses an entire weekly archive and the evaluation in Section~\ref{sec:evaluation} reports automatic measures: corpus composition, status-derivation coverage, scoring sensitivity, and structured-output quality. These are deliberately modest. They confirm that ingestion, scoring, and packet generation behave consistently at archive scale, but they do not validate expiry accuracy against authoritative registers, the fidelity of model summaries, pathway quality, or commercialization outcomes. In particular, a single weekly release yields few expired records, and status derivation leaves roughly a quarter of records in an \emph{unknown} bucket, so the current evidence speaks to engineering reliability rather than decision quality. Expert inter-rater assessment of summaries and pathways, register-checked expiry precision, and longitudinal case studies remain future work.

For a production evaluation, we would treat expiry discovery as an information-retrieval task, legal-status reconciliation as a data-quality task, and patent-to-pathway translation as an expert-review task. Minimum evaluation artifacts should include a gold-standard sample checked against national registers, inter-rater assessment of summary fidelity and pathway usefulness, ablation tests for family-risk penalties, and longitudinal follow-up on whether generated pathway packets improve analyst review time.

\section{Conclusion}
\label{sec:conclusion}

This paper presented an AI-enabled framework for discovering expired and lapsing patents, detecting domain trends, and translating patent disclosures into entrepreneurship-oriented business pathways. By treating patent expiry as a business signal, legal-status input, and archival transition, we connect patent analytics, innovation archive theory, and responsible commercialization support.

The proposed architecture uses tiered ingestion from US, Canadian, and international sources; a legal-status engine; an NLP pipeline; pathway scoring; and four workflow modules backed by auditable Python patterns. Rather than relying on weak illustrative charts or tables, the paper emphasizes how archival records can be converted into structured review packets that combine patent status, pathway hypotheses, compliance or contract evidence, and unresolved legal and market questions.

Future work includes live integration with USPTO ODP and CIPO bulk feeds, expert evaluation of summary fidelity, cross-border family-risk benchmarking, and longitudinal case studies with entrepreneurs and technology transfer offices. We hope this framework helps researchers and practitioners view patent databases not only as legal search systems but as dynamic innovation archives that AI can help activate, with explicit attention to uncertainty, equity, and expert review.

\bibliographystyle{plainnat}
\bibliography{bib/references}

\appendix
\section{Local LLM Pathway Example}
\label{app:ollama}

This appendix reproduces one complete drafting-layer exchange from the real-data proof of concept. The record is CA2616010, ``Medication Compliance System and Associated Methods,'' a granted CIPO patent whose twenty-year term estimate places it about eleven days from expiry as of the reference date, making it a genuine expiry-discovery candidate. The prompt is constructed only from parsed patent text and instructs the model to return strict JSON without asserting legal clearance. The response was produced by Qwen3.6 through a local Ollama instance with sampling temperature 0.1, JSON-constrained output, and hybrid reasoning disabled.

\begin{lstlisting}[caption={Prompt sent to the local model (abstract and claims truncated for space).},label={lst:ollama-prompt}]
Return strict JSON with keys technical_summary, candidate_pathways,
required_evidence, legal_cautions, and reviewer_questions.
Use only the supplied patent text. Do not claim legal clearance.

Patent: CA2616010
Title: MEDICATION COMPLIANCE SYSTEM AND ASSOCIATED METHODS
Status: Granted and Issued
Estimated legal status: active_or_pending
Abstract: A system for monitoring medication compliance in a patient
includes an electronic pill that includes a drug-transporting device
and an antenna positioned on a surface of the drug-transporting device.
A detector is positionable external a gastrointestinal tract of a
patient for detecting a presence of the antenna ...
Claims excerpt: What is Claimed is: 1. A method for making an electronic
pill adapted for monitoring for medication compliance in a patient ...
\end{lstlisting}

\begin{lstlisting}[caption={Structured JSON returned by Qwen3.6 (lists abbreviated).},label={lst:ollama-response}]
{
  "technical_summary": "The patent describes an electronic pill system
    for monitoring medication compliance. The core component is an
    ingestible medication delivery device equipped with an antenna on
    its exterior surface ... The system includes an external detector
    capable of sensing the presence of the antenna within the patient's
    gastrointestinal tract.",
  "candidate_pathways": [
    "Ingestible telemetry systems for physiological monitoring",
    "Smart pill technologies for adherence tracking",
    "External-to-internal wireless communication devices",
    "Flexible substrate electronics for biomedical applications"
  ],
  "required_evidence": [
    "Claims detailing the specific antenna materials",
    "Method for wrapping the flexible substrate and antenna",
    "Specifications for the external detector",
    "Details on temporary power storage in the pill"
  ],
  "legal_cautions": [
    "Based solely on the provided text; not legal advice or an FTO
      opinion.",
    "Status is 'active_or_pending'; verify maintenance and expiry with
      official records.",
    "Claims are limited to the described method and configuration.",
    "Do not rely on this summary for commercial clearance."
  ],
  "reviewer_questions": [
    "How do the antenna materials affect regulatory classification?",
    "What is the operational range of the external detector?",
    "Does power storage allow reliable data transmission?",
    "How does substrate wrapping affect dissolution rate?"
  ]
}
\end{lstlisting}

Two properties of this exchange support the design choices in the main text. First, the model produced well-formed JSON that populated every required field, so it can be consumed directly by the packet schema without post-hoc repair. Second, and more importantly, the useful content is interpretive rather than authoritative: the summary and pathways aid orientation, but the legal cautions and reviewer questions are exactly the items the framework insists must be verified by a human before any commercial or freedom-to-operate decision. The deterministic fallback produces the same field structure with template text when no local model is available, so downstream tooling behaves identically whether or not the model is present.

\end{document}